\documentclass{INTERSPEECH2023}

\usepackage{amsmath,graphicx,multirow}
\usepackage{mathrsfs}
\usepackage{color}
\usepackage{CJKutf8}
\definecolor{red}{rgb}{1,0,0}
\definecolor{green}{RGB}{34,221,34}
\definecolor{blue}{RGB}{43, 145, 213}

\interspeechcameraready

\title{Accurate and Reliable Confidence Estimation Based on Non-Autoregressive End-to-End Speech Recognition System}
\name{Xian Shi, Haoneng Luo, Zhifu Gao, Shiliang Zhang, Zhijie Yan}
\address{Speech Lab of DAMO Academy, Alibaba Group, China}
\email{\{shixian.shi, haoneng.lhn, zhifu.gzf, sly.zsl, zhijie.yzj\}@alibaba-inc.com}

\begin{document}

\maketitle

\begin{abstract}
Estimating confidence scores for recognition results is a classic task in ASR field and of vital importance for kinds of downstream tasks and training strategies. Previous end-to-end~(E2E) based confidence estimation models~(CEM) predict score sequences of equal length with input transcriptions, leading to unreliable estimation when deletion and insertion errors occur. In this paper we proposed CIF-Aligned confidence estimation model~(CA-CEM) to achieve accurate and reliable confidence estimation based on novel non-autoregressive E2E ASR model - Paraformer. CA-CEM utilizes the modeling character of continuous integrate-and-fire~(CIF) mechanism to generate token-synchronous acoustic embedding, which solves the estimation failure issue above. We measure the quality of estimation with AUC and RMSE in token level and ECE-U - a proposed metrics in utterance level. CA-CEM gains 24\% and 19\% relative reduction on ECE-U and also better AUC and RMSE on two test sets. Furthermore, we conduct analysis to explore the potential of CEM for different ASR related usage.

\end{abstract}
\noindent\textbf{Index Terms}: Confidence estimation, end-to-end ASR, non-autoregressive ASR

\section{Introduction}
\label{sec:intro}

Confidence estimation - predicting the probability of true correctness likelihood, is important for classification models in many aspects~\cite{niculescu2005predicting,naeini2015obtaining,guo2017calibration}, and also essential and widely adopted in automatic speech recognition (ASR) field since decades ago~\cite{wessel2001confidence,jiang2005confidence,yu2011calibration}. In human-computer interaction flow,  dialogue system and machine translation system require confidence score of upstream queries to take different actions; Semi-supervised training and active learning use confidence score to guide the training process~\cite{tur2005combining}; In ASR results rescoring and correction, confidence score plays a role as uncertainty measurement~\cite{evermann2000posterior,futami2021asr,deng2022confidence}. At the age of conventional hybrid ASR system (HMM-DNN with external decoder), lattice expanded by transcription contains frame-level phoneme posterior thus it's natural to calculate reliable confidence score which meets the requirements above~\cite{evermann2000posterior,seigel2011combining,ragni2018confidence}.

\begin{figure}[htbp]
    \centering
    \includegraphics[width=0.45\textwidth]{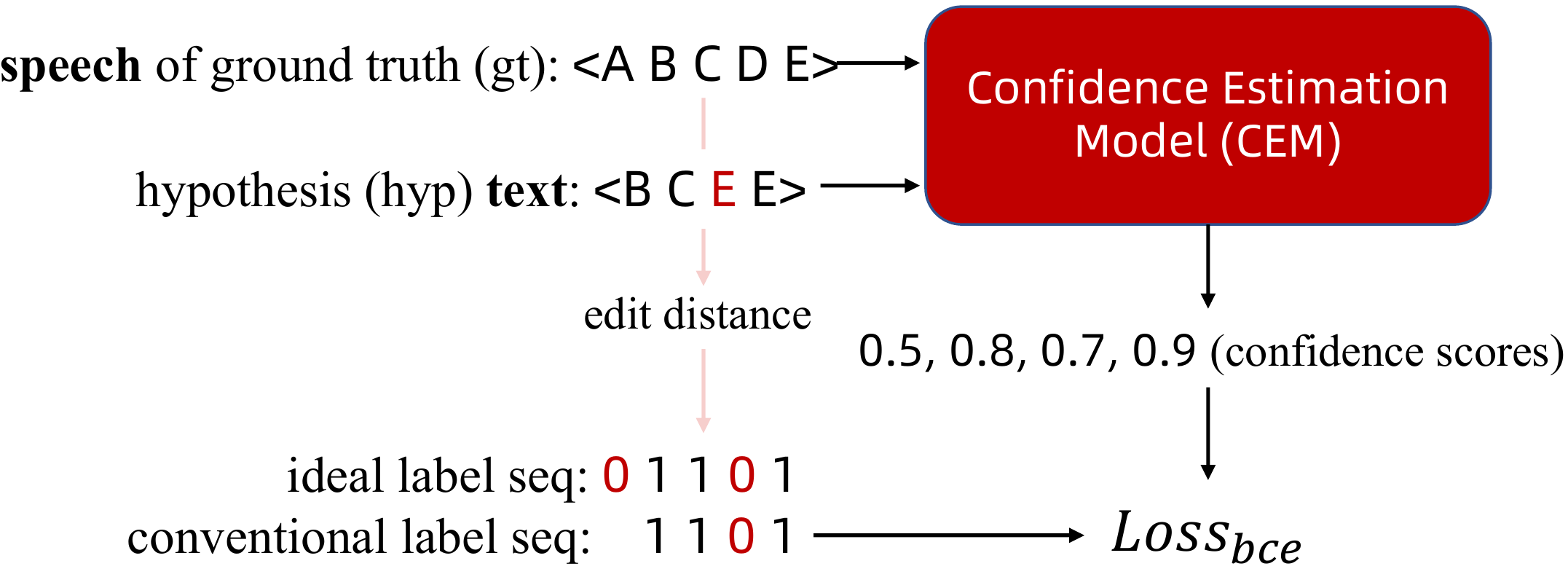}
    \caption{Training of confidence estimation model.}\label{cem_il}
\end{figure}

Recent years have seen dramatic improvement of ASR models in recognition accuracy and various end-to-end (E2E) models predicting the character sequence directly without explicit coupling acoustic model and language model~\cite{vaswani2017attention,kim2017joint,li2019improving,mohamed2019transformers}. 
Meanwhile, the confidence estimation quality collapses as recognition accuracy raises, and \textit{overconfidence} is widely observed in both time-synchronous models and token-synchronous models~\cite{zeyer2021does,chen2016phone,woodward2020confidence}. As discussed in the work of Li et. al.~\cite{li2021confidence}, the posterior given by decoder softmax layer is proved sharp and of poor-quality from the aspect of confidence calibration. 
Previous works make great efforts to address the overconfidence issue and calculate accurate confidence scores for E2E models. In~\cite{futami2021asr}, a sigmoid function is added after BERT~\cite{devlin2019bert} in order to enable the pretrained model to estimate token level confidence scores for further rescoing. A series of works is done by Li et. al. with confidence estimation on attention-based models~\cite{li2021confidence,li2021residual,li2022improving}, placing a fully-connected layer after the E2E decoder is proved an efficient way to perform confidence estimation in listen, attend and spell (LAS) model~\cite{chan2016listen}. The training of estimation module requires hypothesis transcriptions and 0/1 sequences of equal length in pair~(the ideal labels are not available in training) as shown in Fig.~\ref{cem_il}, which leads to a defect that confidence estimation models fail when deletion error occurs. In order to address the defect, \cite{qiu2021multi} introduces deletion prediction task in extra and conducts multi-task training with criterions of different levels.

\begin{figure*}[htbp]
    \centering
    \includegraphics[width=1.0\textwidth]{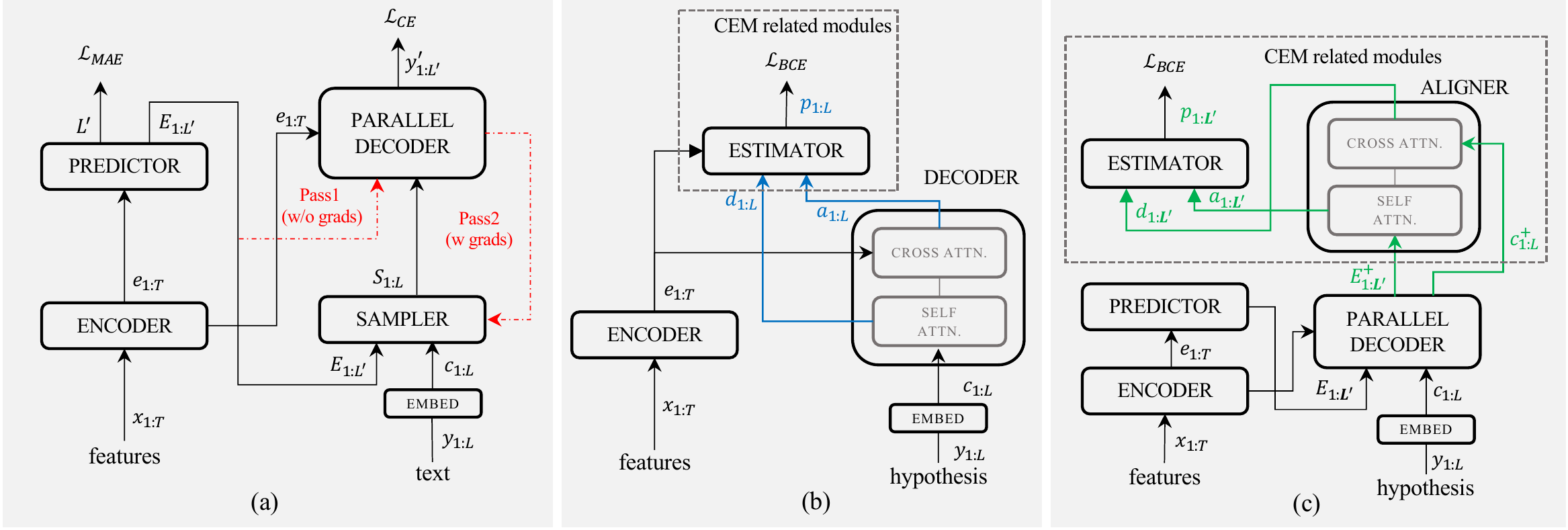}
    \caption{Illustration of the introduced structures. (a) Structure of NAR ASR model Paraformer. (b) Transformer Based Estimator. (c) CIF-Aligned Confidence Estimator.  Dotted boxes indicate the external modules introduced for confidence estimation.}\label{fig1}
    
\end{figure*}
\vspace{-4pt}
Above solutions achieve accurate confidence estimation and deletion prediction exclusively or separately, leaving promotion rooms for a more elegant solution. Paraformer~\cite{gao2022paraformer} is a recently proposed non-autoregressive~(NAR) E2E ASR model. It uses continuous integrate-and-fire~(CIF)~\cite{dong2020cif,yu2021boundary} as predictor to achieve a soft and monotonic alignment mechanism, which transduces encoder output in frames to token-synchronized acoustic embedding. Such structure inspires us a one-shot solution for confidence estimation with natural capacity of deletion prediction.
In this work, we propose CIF-aligned confidence estimator,  which conducts confidence estimation by utilizing the modeling character of CIF predictor in Paraformer. Through an external cross-attention between acoustic embedding and char embedding, CA-CIF predicts confidence score sequence which has the same length with ASR results rather than the given hypothesis, which solves the defect of estimation failure in deletion error cases. We conduct experiments on open source dataset and measure the quality of confidence scores with AUC and ECE-U metrics. The following parts of our paper is organized as below. Section 2 describes the framework of baseline AED~(attention-encoder-decoder) confidence estimator, together with a brief introduction of Paraformer. In Section 3, we represent the proposed CIF-aligned confidence estimator and the metrics we adopt. Section 4 shows the experiments in detail. We make an analysis of the estimators in Section 5 while Section 6 ends the paper with a conclusion.


\vspace{-1mm}
\section{Preliminaries}
\label{sec:format}

\subsection{CIF and Paraformer}

We adopt Paraformer~\cite{gao2022paraformer}, a novel NAR ASR model as our backbone. Briefly, Paraformer achieves non-autoregressive decoding capacity by utilizing CIF~\cite{dong2020cif} and two-pass training strategy (as illustrated in Fig.~\ref{fig1}~(a). A CIF predictor is trained to predict the number of tokens and generate acoustic embedding $\mathbf{E_{1:L^{'}}}$ for parallel decoder, which makes up Pass1 in training (w/o gradient). The char embedding $\mathbf{c_{1:L}}$ will be gradually replaced by $\mathbf{E_{1:L^{'}}}$ as accuracy raises in Pass2. Getting rid of the massive computation overhead introduced by autoregressive decoding and beam-search, Paraformer gains more than 10x speedup with even lower error rate.

\begin{figure}[htbp]
    \centering
    \includegraphics[width=0.35\textwidth]{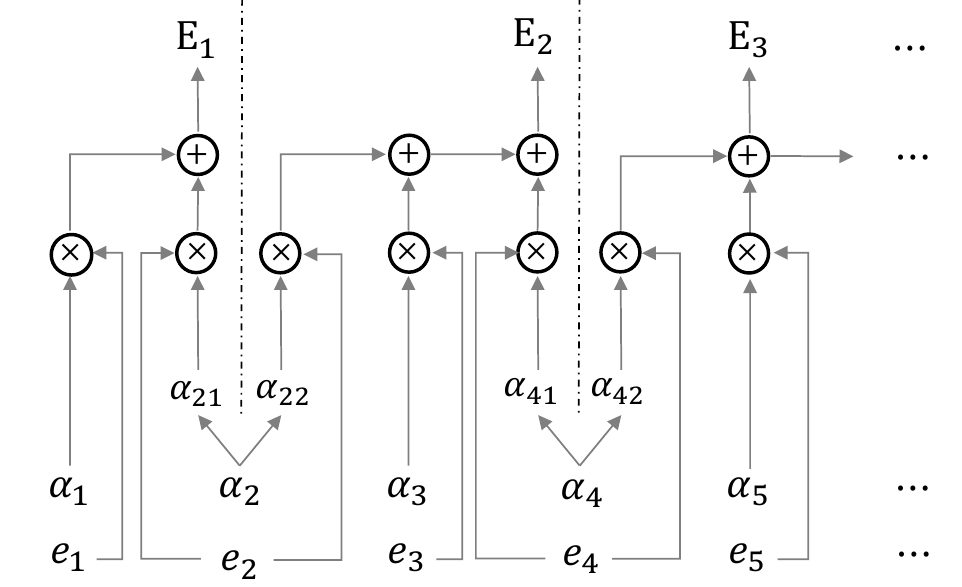}
    \vspace{-3mm}
    \caption{Illustration of integrate-and-fire on encoder output $\mathbf{e}_{1: T}$ with predicted weights $\alpha=(0.3,0.9,0.4,0.4,0.3)$. The integrated acoustic embedding $\mathbf{E}_{1}=0.3\times\mathbf{e}_{1}+0.7\times\mathbf{e}_{2}$, $\mathbf{E}_{2}=0.2\times\mathbf{e}_{2}+0.4\times\mathbf{e}_{3}+0.4\times\mathbf{e}_{4}$. The sum of weights $\alpha$ is $L^{'}$ - the length of prediction sequence.}\label{cif}
\end{figure}

\vspace{-3mm}
\subsection{Transformer Based Estimator}

We implement confidence estimation with Transformer using the similar structure as~\cite{li2021confidence}, which is based on LAS. Considering an utterance features $\mathbf{x}_1,...,\mathbf{x}_T$ and the hypothesis sequence $\mathbf{y}_1,...\mathbf{y}_L$, the acoustic representation and character embedding with the well-trained ASR model parameters is obtained as:
\begin{equation}
\label{eq1}
\begin{aligned}
\mathbf{e}_{1: T} &=\operatorname{ENCODER}\left(\mathbf{x}_{1: T}\right) \\
\mathbf{c}_{1: L} &=\operatorname{EMB}(\mathbf{y}_{1: L}) \\
\end{aligned}
\end{equation}
The output of self-attention $\mathbf{d}_l$ inside decoder is reserved for confidence estimator (EST),
\begin{equation}
\label{eq2}
\begin{aligned}
\mathbf{a}_{1: L}, \mathbf{d}_{1: L} &=\operatorname{DECODER}\left(\mathbf{e}_{1: T},\mathbf{c}_{1: L}\right) \\
p_{1: L} &=\operatorname{sigmoid}\left(\mathrm{EST}\left(\mathbf{a}_{1: L}, \mathbf{d}_{1: L}; \mathbf{e}_{1: T}\right)\right)
\end{aligned}
\end{equation}
The external confidence estimation module composed of several attention layers (same as Transformer decoder) and a sigmoid function predicts the token's confidence score $p_l$. The forward process is conducted in full sequence rather than autoregressive mode. Treating $p_l$ as a posterior probability of the given token being correct, the estimator can be trained under binary cross entropy loss:
\begin{equation}
\mathcal{L}(\mathbf{c}, \mathbf{p})=-\frac{1}{L} \sum_{l=1}^{L}\left(c_{l} \log \left(p_{l}\right)+\left(1-c_{l}\right) \log \left(1-p_{l}\right)\right)
\end{equation}
where $\mathbf{c}$ refers to the label sequence calculated in advance according to edit distance.


\section{Proposed Methods}
\subsection{CIF-Aligned Estimator}


Among the candidate inputs of Transformer based estimator (eq~\ref{eq2}), $\textbf{a}_{1: L}$ obtained by cross-attention is of vital importance. It contains the soft alignment information of acoustic representation and character embedding, which is the key of confidence estimation but also the culprit of the unreliable prediction: label sequence should be synchronous with hypothesis tokens. An ideal estimator is supposed to predict confidence scores with the same length as ground truth regardless of hypothesis length. 

In Paraformer, the predictor module conducts CIF over encoder outputs and obtains acoustic embedding, which meets the requirements above exactly: carrying acoustic information with equal length as recognition results. 
In light of such capability, we propose CIF-aligned estimator: adopting an cross attention module~(named Aligner) between $\mathbf{E}_{1: L^{'}}$ and $\mathbf{c}_{1: L}$ to calculate attention matrix synchronous with $\mathbf{E}_{1: L^{'}}$. Formally, $p_{1: L^{'}}$ is calculated as follow:
\begin{equation}
\begin{aligned}
\mathbf{E}_{1: L^{'}} &=\operatorname{CIF}\left(\mathbf{e}_{1: T}\right) \\
\mathbf{a}_{1: L^{'}}, \mathbf{d}_{1: L^{'}} &=\operatorname{ALIGNER}\left(\mathbf{c}_{1: L}, \mathbf{E}_{1: L^{'}}\right) \\
p_{1: L^{'}} &=\operatorname{sigmoid}\left(\mathrm{EST}\left(\mathbf{a}_{1: L^{'}}, \mathbf{d}_{1: L^{'}} \right)\right)
\end{aligned}
\end{equation}
$L^{'}$ in the expressions indicates the sequence is of same length as CIF prediction rather than the hypothesis.
In experiments, however, we find it difficult for a randomly initialized decoder to establish contact between $\mathbf{c}_{1: L}$ and $\mathbf{E}_{1: L^{'}}$ directly, they are thus replaced by their encoder-attended representations of parallel decoder outputs:
\begin{equation}
\begin{aligned}
\begin{split}
\mathbf{c}^{+}_{1: L} &=\operatorname{DECODER}\left(\mathbf{e}_{1: T},\mathbf{c}_{1: L}\right) \\
\mathbf{E}^{+}_{1: L^{'}} &=\operatorname{DECODER}\left(\mathbf{e}_{1: T},\mathbf{E}_{1: L^{'}}\right) \\
\end{split}\\
\begin{split}
\mathbf{a}_{1: L^{'}}, \mathbf{d}_{1: L^{'}} &=\operatorname{ALIGNER}\left(\mathbf{c}^{+}_{1: L}, \mathbf{E}^{+}_{1: L^{'}}\right) \\
\end{split}\\
\end{aligned}
\end{equation}
Parallel decoder takes both $\mathbf{c}_{1: L}$ and $\mathbf{E}_{1: L^{'}}$ might look confusing but it turns out that through the two-pass training strategy of Paraformer, the decoder attention maintains the capacity of processing both acoustic embedding and char embedding. CIF-Aligned estimator is exactly and naturally based on such capacity - an external attention aligner is introduced to conduct cross-attention over the embeddings and embed char embedding into length of $L^{'}$. Since the label sequence length for CA-CEM should be $L^{'}$, in practice we use the recognition results of base ASR model to conduct edit-distance alignment with hypothesis sequences to generate true or false sequence.

\subsection{Evaluation Metrics}
In binary classification tasks, the area under the curve (AUC) is always chosen to be one of the evaluation metrics, where curve refers to receiver operating characteristic (ROC) curve. For posterior probability predicted by an estimator, selecting threshold $\Tilde{p}$ with equal spacing in $[0, 1]$ then calculate true positive rate (TPR) and false positive rate (FPR) with $\Tilde{p}$:
\begin{equation}
\operatorname{TPR}(\tilde{p})=\frac{\mathrm{TP}(\tilde{p})}{\operatorname{TP}(\tilde{p})+\operatorname{FN}(\tilde{p})},
\operatorname{FPR}(\tilde{p})=\frac{\mathrm{FP}(\tilde{p})}{\operatorname{FP}(\tilde{p})+\operatorname{TN}(\tilde{p})}\nonumber
\end{equation}
The values of TPRs and FPRs make up an ROC curve and AUC is thus calculated. AUC usually ranges in $[0.5, 1.0]$, higher value indicates a better estimator.

Previous works have also seen normalized cross entropy (NCE) and equal error rate (EER) adopted for ASR confidence estimation evaluation~\cite{li2021residual,li2022improving}. These metrics, however, evaluate the estimators from a similar perspective - quality of token level confidence score. In this work, we propose to use \textit{expected calibration error - utterance level} (ECE-U) as another metrics. ECE is not an appropriate index for binary classification tasks as the accuracy is relevant to acceptance threshold. But in utterance level, it exactly reflects the distance between averaged token confidence score and CER. ECE is calculated by partitioning predictions into $M$ equally-spaced bins and taking a weighted average of the bin's accuracy/confidence difference, formally:
\begin{equation}
\sum_{m=1}^{M} \frac{\left|B_{m}\right|}{n}\left|\operatorname{acc}\left(B_{m}\right)-\operatorname{conf}\left(B_{m}\right)\right|\nonumber
\end{equation}
ECE-U measure the distance in utterance level, which is more appropriate for ASR:
\begin{equation}
\sum_{m=1}^{M} \frac{\left|B_{m}\right|}{n}\left(\sum_{u\in{B_{m}}} \left|1-\operatorname{CER}\left(hyp_u, gt_u\right)-\operatorname{conf}\left(u\right)\right|\right)\nonumber
\end{equation}

\section{Experiments}

\label{sec:typestyle}
\subsection{Data Introduction}
Two data sets are used in our experiments.
Aishell-1 is an open-source ASR data set containing 178 hours Mandarin speech, which is a mostly chosen testbed to verify ASR model performance. We prepare the data for confidence estimator training in advance: For a given hypothesis, we generate the true or false sequence by calculating the edit distance comparing with ground truth~(for AED-CEM and NAR-CEM) or decoding results of ASR base model~(for CA-CEM). In order to avoid the unbalance of true/false labels, we use the model before complete convergence for decoding~(CER over train set is 13.6\%). With the same strategy, two dev/test sets are prepared with CER of 15.3/16.9 and 33.5/35.9, named \textit{Test15} and \textit{Test30} respectively. Such data reflects the typical use case of confidence score like self-supervised training and oral evaluation where the CER is moderately high. In industrial experiment stage, we use a 20,000-hour Mandarin corpus as train set and meeting scenario corpus of real scene as test sets.

\vspace{-2mm}
\subsection{Experimental Setup}
We conduct the confidence estimation experiments with the open-source ESPnet toolkit~\cite{watanabe2018espnet}, the networks are implemented in PyTorch. In Aishell-1 experiments, well-trained Conformer and Paraformer (12-layer encoder and 6-layer decoder, trained for 50 epochs under noam learning rate schedule) are used to initialize confidence estimation networks' parameters. Following the architecture proposed by~\cite{li2021confidence}, we reproduce the AED-CEM in Mandarin corpus as baseline. In the training process, encoder is freezing thus the external parameter introduced by confidence estimation contains decoder, estimator and aligner. All of the estimation networks are trained with exactly same configurations: training for 128,000 steps, warm-up for 10,000 steps with dropout 0.15 for the entire network. Using a single TESLA V100 GPU, training process of a confidence estimation model cost around 30 hours.

\begin{table*}[htbp]
\centering
\caption{Results of the confidence estimations.}\label{res}
\vspace{-2mm}
\begin{tabular}{llllllll}
\hline
\multirow{2}{*}{}                                     & \multicolumn{3}{c}{Test15~(dev/test)}                                  &  & \multicolumn{3}{c}{Test30~(dev/test)}                                  \\ \cline{2-4} \cline{6-8} 
                                                      & AUC~$\uparrow$  & ECE-U(\%)~$\downarrow$ & RMSE~$\downarrow$ &  & AUC~$\uparrow$  & ECE-U(\%)~$\downarrow$ & RMSE~$\downarrow$ \\ \cline{1-4} \cline{6-8} 
Conformer-Softmax                                     & 0.884/0.877          & 7.63/7.18                   & 0.120/0.121            &  & 0.804/0.801          & 8.93/8.76                   & 0.156/0.159            \\
Paraformer-Softmax                                    & 0.895/0.892          & 7.48/7.08                   & 0.117/0.118            &  & 0.834/0.827          & 9.29/8.66                   & 0.151/0.152            \\ \cline{1-4} \cline{6-8} 
AED-CEM~\cite{li2021confidence} & 0.953/0.951          & 2.27/2.52                   & 0.068/0.074            &  & 0.965/0.964          & 4.43/4.83                   & 0.080/0.084            \\
NAR-CEM                                               & \textbf{0.957/0.955} & 2.10/2.38                   & 0.067/0.073            &  & 0.967/0.966          & 4.41/4.77                   & 0.079/0.084            \\
CA-CEM                                                & \textbf{0.958/0.956} & \textbf{1.79/1.90}          & \textbf{0.063/0.067}   &  & \textbf{0.970/0.968} & \textbf{3.37/4.00}          & \textbf{0.068/0.074}   \\ \hline
\end{tabular}
\end{table*}

\vspace{-2mm}
\subsection{Results}
\textbf{Comparing softmax and the three CEM models.}
Confidence estimation with AED based estimator is our baseline, and the logits probability given by decoder softmax is also shown for comparison. NAR-CEM refers to implementing the same estimator as AED-CEM inside Paraformer. Intuitively, we also calculate root-mean-square deviation~(RMSE) between averaging confidence and accuracy. The experiments results shown in Table~\ref{res} proves that: (i) External estimator greatly outperforms using logits as confidence, the token level confidence is of better quality and utterance level confidence is closer to accuracy; (ii) Comparing AED-CEM and NAR-CEM, the later one slightly benefits from the NAR decoder - it models char embedding with global context; (iii) CA-CEM achieves the best performance in the table, the AUC is equal to NAR-CEM but ECE-U descends 15\% to 23\% relatively, which also indicates that measuring the performance of estimation with only AUC is biased - estimators with similar AUC may differ a lot in utterance level confidence quality; (iv) Comparing models' performance on \textit{Test15} and \textit{Test30}, higher CER leads to a more difficult task to approximating the accuracy with confidence.
It is noteworthy that the training process of CEM models is sensitive to several hyper-parameters, especially dropout for aligner attention and warm-up learning rate schedule. The table below shows the confidence scores of the three confidence models~(from Aishell-1 testset, Softmax conf. is provided by a Conformer model in early steps, for general reference only). We find that CA-CEM is able to figure out the deletion error position and achieves closer average confidence to the real accuracy.
\vspace{-2mm}
\begin{table}[htbp]
\caption{Confidence scores of models in a case from Aishell-1}
\vspace{-2mm}
\resizebox{.99\columnwidth}{!}{
\begin{tabular}{lll}
\hline
GroundTruth    & \begin{CJK*}{UTF8}{gbsn}~美~~~~国~~~~东~~~~卡~~~~罗~~~~莱~~~~纳~~~~大~~~~学\end{CJK*}                                    & acc=77.8\% \\ \cline{3-3} 
Hypothesis     & \begin{CJK*}{UTF8}{gbsn}~\textcolor{red}{北}~~~~国~~~~东~~~~卡~~~~罗~~~~莱~~~~\textcolor{red}{**}~~~~大~~~~学\end{CJK*}                                   & Avg. conf. \\ \hline
Softmax conf.* & 0.90 0.94 0.82 0.74 0.79 0.52 ~**~~  0.73 0.98 & 0.803      \\ \hline
AED-CEM conf.  & \textcolor{blue}{0.04} 0.99 0.99 0.99 0.98 0.94 ~**~~  0.99 0.99 & 0.864      \\
CA-CEM conf.   & \textcolor{green}{0.07} 0.99 0.96 0.94 0.99 0.98 \textcolor{green}{0.05} 0.98 0.99 & \textbf{0.772}      \\ \hline
 \end{tabular}}
\end{table}
\vspace{-2mm}
\vspace{-1mm}
\section{Analysis}


\noindent\textbf{Confidence under different acoustic environments.} 
An ideal CEM is supposed to be able to measure the clarity of speech.
Table~\ref{alim} gives us more insight into how confidence score varies under different acoustic environments. \textit{TestMeeting} is a meeting scenario corpus of real scene with 2400 utterances, and we prepare three augmented forms of it: adding noise or reverberation, processing the noise form with a speech enhancement front-end~(FE) model. With a CA-CEM trained with 20,000-hour industrial dataset, we conduct confidence estimation with both ground truth and hypothesis transcriptions. The results show that as the audio quality decreases (CER increase), the averaging confidence score also declines. Such performance indicates that confidence estimation model is also capable to measure the speech quality from the perspective of ASR~(rather than human sense) without ground truth transcriptions.
\vspace{-2mm}
\begin{table}[ht]
\centering
\caption{Confidence in different acoustic environments.}\label{alim}
\vspace{-2mm}
\begin{tabular}{llll}
\hline
\multirow{2}{*}{} & \multirow{2}{*}{CER(\%)} & \multicolumn{2}{c}{avg. CEM conf.} \\ \cline{3-4} 
                  &                          & GT trans        & HYP trans        \\ \hline
Clean             & 8.46                     & 0.914           & 0.932            \\
Noise            & 17.24                    & 0.860           & 0.904            \\
RIR              & 17.73                    & 0.855           & 0.893            \\
Noise+FE                & 24.70                    & 0.790           & 0.857            \\ \hline
\end{tabular}
\end{table}

\begin{figure}[htb]
    \centering
    \includegraphics[width=0.43\textwidth]{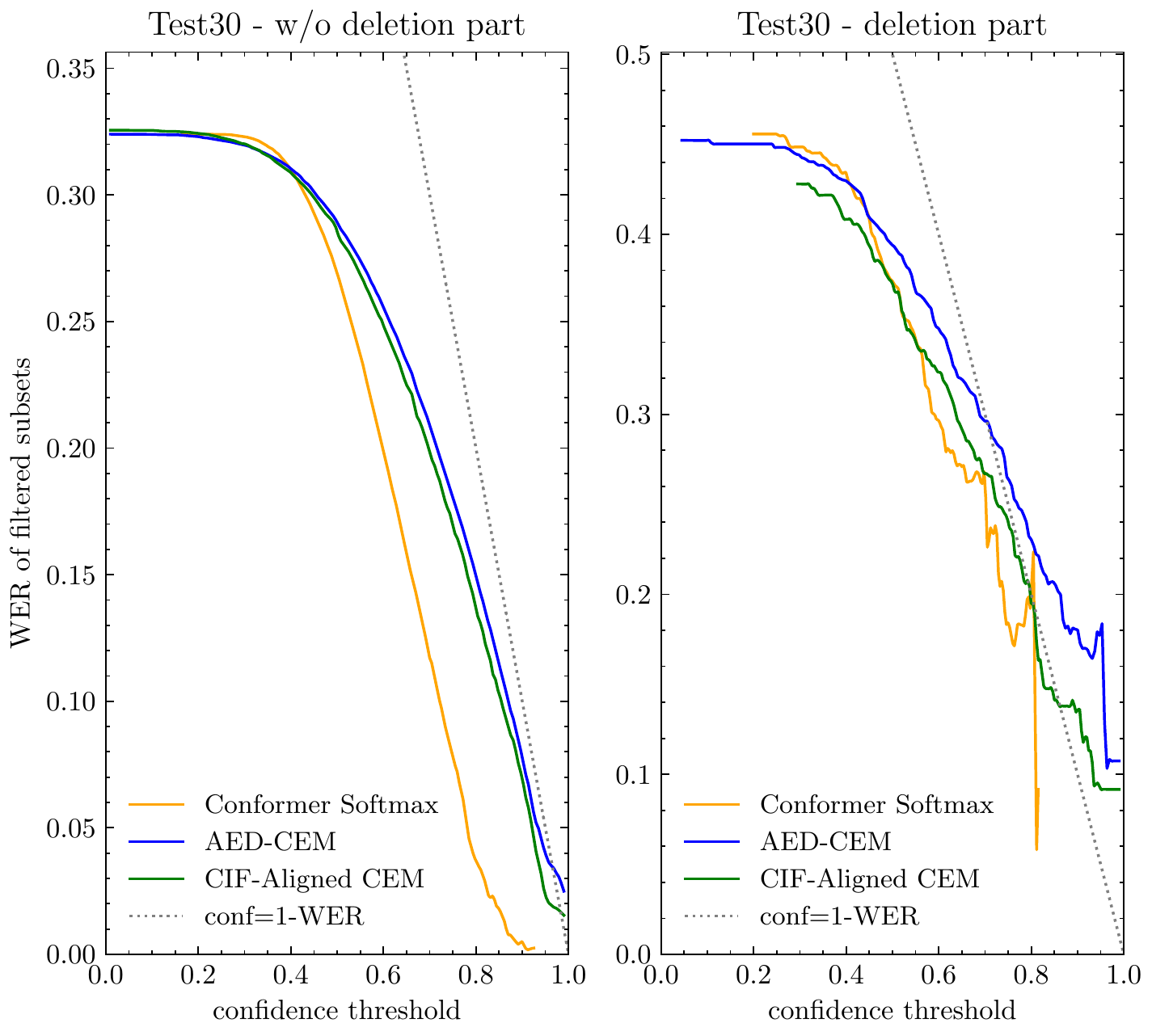}
    \caption{Filtered WER - threshold curves over \textit{Test30}.}
    \label{figf}
\end{figure}

\noindent\textbf{Unlabeled data selection with confidence estimation.} 
Selecting unlabeled speech data with confidence estimation module is a widely adopted strategy. For ASR systems in continuous training iteration, discovering data of poor performance can reduce training overhead and improve efficiency. Moreover, CEM is also able to measure the distance between models and datasets in the aspects of acoustic and linguistic. Conducting such selection requires the linear and monotonic correlation between utterance level confidence score and CER. We filter the subsets with confidence higher than thresholds on x-axis, then plot the subset WER and filtering threshold curves of the three estimations in Fig.~\ref{figf}. The same plotting is conducted in two parts of \textit{Test30}: Utterance without deletion error~(left) and with deletion error~(right). 
It turns out that in the cases of insertion/substitution errors, softmax and confidence estimators both achieve a linear and monotonic correlation. Sharp spikes are observed in curves of deletion part, which is similiar with~\cite{li2021confidence}. The internal causes in our experiments might not be \textit{overconfidence} as discussed in~\cite{li2021confidence}. Actually, as the threshold increases, the filtered subsets are getting smaller. Utterance with high confidence score but low accuracy will directly leads to spikes in the right of the curve, and deletion error is always the culprit. Obviously the entire curve is influenced by the unreliable estimation of deletion from softmax and AED-CEM, but CA-CEM overcomes the shortcoming with the cross-attention in aligner.

\vspace{-5mm}

\section{Conclusion}
\label{ssec:subhead}

In this paper, we conduct confidence estimation module on a novel non-autoregressive end-to-end ASR model - Paraformer. Making good use of the modeling character of CIF predictor, the proposed CIF-Aligned confidence estimator achieves accurate and reliable estimations which outperforms the previous AED-based confidence estimator, and also overcomes the disadvantage of estimation failure in deletion errors. We measure the confidence scores in two aspects with the following metrics: AUC in token level and ECE-U/RMSE in utterance level. We analyse the performance of the estimations in different acoustic environments, which turns out that confidence estimation might be used in evaluating the quality of speech. Moreover, we discuss the inner cause of the spike issue in unlabeled data selection, and prove the effectiveness of the proposed estimator.

\bibliographystyle{IEEEtran}
\bibliography{mybib}

\end{document}